# Numerical scheme for nonlinear optical response of metallic nanostructure: Quantum hydrodynamic theory solved by adopting effective Schrödinger equation

**Takashi Takeuchi[1,*] and Kazuhiro Yabana[1,**]**

[1] Center for Computational Sciences, University of Tsukuba, Tsukuba 305–8577, Japan

[*]Email: take@ccs.tsukuba.ac.jp; [**]Email: yabana@nucl.ph.tsukuba.ac.jp;

**Abstract** – Quantum hydrodynamic theory (QHT) can describe some of the characteristic features of quantum electron dynamics that appear in metallic nanostructures, such as spatial nonlocality, electron spill-out, and quantum tunneling. Furthermore, numerical simulations based on QHT are more efficient than fully quantum mechanical approaches, as exemplified by time-dependent density functional theory using a jellium model. However, QHT involves kinetic energy functionals, the practical implementation of which typically induces significant numerical instabilities, particularly in nonlinear optical phenomena. To mitigate this problem, we develop a numerical solution to QHT that is quite stable, even in a nonlinear regime. The key to our approach is to rewrite the dynamical equation of QHT using the effective Schrödinger equation. We apply the new method to the linear and nonlinear responses of a metallic nanoparticle and compare the results with fully quantum mechanical calculations. The results demonstrate the numerical stability of our method, as well as the reliability and limitations of QHT.

**I. INTRODUCTION**

Metallic nanoparticles were generated in a vacuum in the mid-1980s, and since then, their structural and optical properties have been extensively investigated [1]. The quantum mechanical effects in metallic nanoparticles and their size dependence have been explored theoretically [2]. Subsequently, metallic nanostructures fabricated on a substrate became a typical platform for plasmonic systems, which enabled a new way to delicately control the light-matter interactions utilized in various applications, such as optoelectronic devices, solar cells, and biosensors [3-6]. Recently, quantum mechanical effects in the optical response of nanostructures have attracted renewed interest and have been demonstrated both theoretically and experimentally. Specifically, when the size of the nanostructures is as small as several nanometers, the optical response is affected by spatial nonlocality [7-13] and electron spill-out [14-19]. Furthermore, state-of-the-art fabrication techniques have enabled nanostructures with angstrom-level fineness separated by subnanometer gaps between plasmonic particles [20-25]. At such an extremely microscopic scale, electrons in the gaps exhibit quantum tunneling [20-22, 26-33]. These quantum effects have attracted attention for enhancing optical nonlinearity [8, 19, 31, 33], which is important for developing and improving a wide range of applications [34-39]. In this context, there is a growing interest in establishing numerical schemes to analyze the optical nonlinearity caused by quantum effects in metallic nanostructures.

The most straightforward way to accommodate this demand is via a first-principles calculation that employs electronic orbitals, such as time-dependent density functional theory (TDDFT) [40, 41]. Although the feasibility of first-principles calculations is limited to small systems, using the TDDFT with a jellium model (JM) that can capture important features of the quantum effects may be applicable to nanosystems comprising up to a few thousand electrons [2]. Therefore, the combination of TDDFT and the JM has been extensively used in the last decade to analyze the optical properties of isolated nanodimers (two metallic nanoparticles) with a subnanometer gap, and this combination could accurately reproduce experimental results that include electron tunneling through the gap [20, 21]. However, numerical calculations based on TDDFT with the JM for three-dimensional nanostructures is significantly time-consuming because its computational cost is proportional to the product of the spatial size of the system and the number of orbitals. Therefore, even if the calculation is executed on cutting-edge supercomputers, the nanostructure length that can be calculated within a reasonable computational time is limited to approximately 10 nm. However, the typical size of nanostructures in real measurements is considerably larger—several tens or hundreds of nanometers in most cases [3-6].

A semiclassical approach using kinetic energy (KE) functionals was developed to overcome the computational restrictions of the TDDFT with the JM, and it has recently gained interest in the field of nanoplasmonics. In this approach, the electron dynamics is treated as a fluid, the motion of which is described by local physical quantities, such as the electron density $n(\mathbf{r}, t)$ and electric current density $\mathbf{J}(\mathbf{r}, t)$ or velocity field $\mathbf{v}(\mathbf{r}, t)$. The semiclassical approach no longer involves electronic orbitals; therefore, it enables calculations that include large nanostructures with an acceptable computational cost. Such an approach with KE functionals has a long history in various fields [42-45]. In the last decade, this approach has attracted attention in the field of nanoplasmonics, taking into account the quantum mechanical effects. Specifically, the spatial nonlinearity of nanostructures was first described using the Thomas-Fermi (TF) KE functional, which depends on $n$ [7-12]. Subsequently, the von Weizsäcker (vW) KE functional, which has a $\nabla n$ dependence, has been incorporated to address electron spill-out and quantum tunneling [14-19]. This is referred to as TF$\lambda$vW, and it includes a phenomenological parameter $\lambda$, the value of which is adjusted to reproduce the electron spill-out in the TDDFT calculation. Very recently, higher-order KE functionals, including the $\nabla^2 n$ contribution, have been introduced to more precisely describe electron dynamics [46]. In previous literature, semiclassical approaches with and beyond TF$\lambda$vW are referred to as quantum hydrodynamic theory (QHT). Notably, in one study, QHT was successfully derived from TDDFT, which clarifies their mutual correspondence [18].

QHT is a promising avenue to analyze the quantum mechanical effects in the optical response of metallic nanostructures; however, numerical calculations based on QHT have thus far been limited to linear optical responses, and they are hardly applied to nonlinear optical phenomena. This is owing to the computational difficulty of the KE functionals, in which $1/n$ and/or the cube root of $n$ are involved. Therefore, even if the density becomes zero or negative ($n \leq 0$) at a single spatial point, owing to numerical errors in the update of $n$, the QHT calculation immediately breaks down. It was difficult to develop a numerical scheme that exactly assures $n > 0$ in the whole spatial region and every time step; therefore, almost all previous studies using QHT have focused on the linear response of metallic nanostructures, where a linear equation for the change in the density from the ground state is calculated, whereas the update of the density is unnecessary [14-18]. To the best of our knowledge, only one very recent study has addressed the nonlinear optical response using QHT with TF$\lambda$vW (called QHT$-$TF$\lambda$vW hereafter), and this study revealed the mechanism of second harmonic generation originating from the electron spill-out at the interface of a metallic thin film [19]. However, in their numerical implementation, a perturbative expansion was adopted to describe the second-order optical nonlinearity. Therefore, it is necessary to develop a new code to analyze higher-order nonlinearities that involve growing complexity as the order of the nonlinearity increases. Furthermore, owing to the perturbative treatment, their calculation was compelled to adopt an undepleted pump approximation, in which the fundamental field is not affected by the nonlinear process; however, it becomes important in some nonlinear optical applications [34-36].

In the present study, a numerical scheme is developed to stably calculate both linear and nonlinear electron dynamics based on QHT−TF$\lambda$vW. The key component of our approach is to rewrite the QHT equation using the effective Schrödinger equation (ESE), which ensures a stable calculation of the KE functionals without any numerical instabilities. This ESE involves only one orbital; therefore, the computational cost is considerably lower than that of the TDDFT with the JM. The time evolution of the orbital can be solved using, for example, the real-time and -space method, which has been well tested in previous studies. As a demonstration, we present the calculations of the linear and third-order nonlinear optical responses of a metallic nanosphere, and we compare the results to those of the TDDFT with the JM. We show that the two calculations are in reasonable agreement with each other.

The remainder of this paper is organized as follows. In Sec. II, we introduce QHT and rewrite the equation using the ESE. A coupled optical and electronic system is also described in the QHT using the ESE. In Sec. III, the calculated results for nanospheres are presented and compared with those obtained using DFT with the JM for the ground state and those obtained using TDDFT with the JM for the linear and nonlinear responses. Finally, the conclusions are presented in Sec. IV.

## II. THEORETICAL FRAMEWORK

### A. Summary of TDDFT and QHT formalisms

In this subsection, we present the basic formulas of TDDFT and QHT for which the numerical methods are developed. A direct derivation of QHT from TDDFT is presented based on a description given in the literature [18].

We consider an isolated system, in which a metallic nanostructure is subjected to an electromagnetic field. In TDDFT, the electron dynamics is described by the following time-dependent Kohn-Sham (KS) equation for electronic orbitals $\psi_j(\mathbf{r}, t)$:

$$i\hbar \frac{\partial \psi_j}{\partial t} = \left[ \frac{\{-i\hbar\nabla + e(\mathbf{A} + \mathbf{A}_{\text{XC}})\}^2}{2m} - e\phi + \frac{\delta E_{\text{XC}}}{\delta n} \right] \psi_j, \tag{1}$$

where $\mathbf{A}(\mathbf{r}, t)$ and $\phi(\mathbf{r}, t)$ are the vector- and scalar-potentials, respectively, which are related to the electric and magnetic fields $\mathbf{E}(\mathbf{r}, t)$ and $\mathbf{B}(\mathbf{r}, t)$ as $\mathbf{E}(\mathbf{r}, t) = -\partial \mathbf{A}/\partial t - \nabla\phi$ and $\mathbf{B} = \nabla \times \mathbf{A}$. $\mathbf{A}_{\text{XC}}(\mathbf{r}, t)$ and $\delta E_{\text{XC}}/\delta n = V_{\text{XC}}(\mathbf{r}, t)$ are the exchange-correlation (XC) vector and scalar potentials, respectively, and $\hbar$, $m$, and $e$ are the reduced Planck constant, electron mass, and elementary charge, respectively. Throughout this study, we adopt SI units. The electron density $n(\mathbf{r}, t)$ and electric current density $\mathbf{J}(\mathbf{r}, t)$ are calculated from $\psi_j$ as

$$n = \sum_{j}^{\text{occ}} |\psi_j|^2, \tag{2}$$

$$\mathbf{J} = -\frac{e}{m} \text{Re} \left[ \sum_{j}^{\text{occ}} \psi_j^* \{-i\hbar\nabla + e(\mathbf{A} + \mathbf{A}_{\text{XC}})\} \psi_j \right]. \tag{3}$$

The time evolutions of $\mathbf{A}$ and $\phi$ are governed by the following inhomogeneous wave equations with $n$ and $\mathbf{J}$:

$$\frac{\partial \nabla \cdot \mathbf{A}}{\partial t} + \nabla^2 \phi = -\frac{e(n^{(+)} - n)}{\epsilon_0}, \tag{4}$$

$$\frac{\partial^2 \mathbf{A}}{\partial t^2} + \frac{\partial \nabla \phi}{\partial t} + \frac{1}{\epsilon_0 \mu_0} \nabla \times \nabla \times \mathbf{A} = \frac{\mathbf{J}}{\epsilon_0}, \tag{5}$$

where $n^{(+)}(\mathbf{r})$ is the external positive density, such as those from ions, and $\epsilon_0$ and $\mu_0$ are the permittivity and permeability of free space, respectively. In TDDFT targeted at an isolated system, such as an atom, molecule, or nanostructure, Eqs. (1–4) are numerically solved by adopting the length gauge, and the dipole approximation is imposed.

In QHT, we describe the electron dynamics using the equation for the electric current density $\mathbf{J}(\mathbf{r}, t)$. In the case of QHT−TF$\lambda$vW, which is used in this study, it is given by [18]

$$\frac{\partial \mathbf{J}}{\partial t} = \frac{ne^2}{m}\mathbf{E} - \frac{e}{m}\mathbf{J} \times (\mathbf{B} + \nabla \times \mathbf{A}_{\mathrm{XC}}) + \frac{1}{e}\left\{\frac{\mathbf{J}}{n}(\nabla \cdot \mathbf{J}) + (\mathbf{J} \cdot \nabla)\frac{\mathbf{J}}{n}\right\}$$

$$+ \frac{ne}{m}\nabla\left(\frac{\delta E_{\mathrm{XC}}}{\delta n} + \frac{\delta T_{\mathrm{TF}}}{\delta n} + \lambda\frac{\delta T_w}{\delta n}\right). \tag{6}$$

Here, the first and second terms on the right-hand side are from the electromagnetic fields, $\mathbf{E}$ and $\mathbf{B}$, respectively, which are equivalent to the Lorentz force. The third term, which is composed of $n$ and $\mathbf{J}$, is the convective term, and the fourth term comes from the energy functionals. $\mathbf{A}_{\mathrm{XC}}(\mathbf{r}, t)$ and $\delta E_{\mathrm{XC}}/\delta n = V_{\mathrm{XC}}(\mathbf{r}, t)$ are the XC vector and scalar potentials, respectively, which have already appeared in the TDKS equation, given in Eq. (1). $T_{\mathrm{TF}}$ is the TF KE functional, and its functional derivative is given as:

$$\frac{\delta T_{\mathrm{TF}}}{\delta n} = \frac{5}{3}c_{\mathrm{TF}}n^{2/3}, \tag{7}$$

where $c_{\mathrm{TF}} = \frac{\hbar^2}{m}\frac{3}{10}(3\pi)^{2/3}$. $\delta T_w/\delta n$ is the functional derivative of the vW KE functional:

$$\frac{\delta T_{\mathrm{w}}}{\delta n} = \frac{\hbar^2}{8m}\left(\frac{\nabla n \cdot \nabla n}{n^2} - 2\frac{\nabla^2 n}{n}\right).\tag{8}$$

In Eq. (6), an adjustable phenomenological parameter $\lambda$ is included. This value depends on the derivation of Eq. (6), which ranges from $1/9$ to $1$ [45]. We will later determine the value from the condition that the electron spill-out is reasonably described.

The history of the fluid dynamical description of fermionic many-body systems originates from the TF theory in 1927 [42, 43]. The vW correction term in Eq. (8) was first derived in 1935 [44], which showed that $\lambda = 1$. Then, an expression with $\lambda = 1/9$ was derived using the density expansion method [45]. Notably, Ciraci recently provided a direct derivation of Eq. (6) from the TDDFT [18]. They have also provided high-order KE functionals systematically [46]; however, these are beyond the scope of the present study.

The linear optical response based on QHT$-$TF$\lambda$vW can be executed by linearizing Eqs. (6–8), where it is unnecessary to explicitly calculate the time evolution of electron density. However, to extend the application of this method to a nonlinear regime, an evolution of the electron density $n(\mathbf{r}, t)$ is required. This is typically achieved by numerically employing the equation of continuity, $\frac{\partial n}{\partial t} = \frac{1}{e}\nabla \cdot \mathbf{J}$. However, the time evolution calculation of the electron density is often accompanied by numerical difficulty. Density is a positive definite quantity; however, it is not easy to implement a numerical method that strictly ensures it. Additionally, an accurate calculation of the density is difficult in spatial regions where the density is extremely low. The $\delta T_{\mathrm{w}}/\delta n$ term includes $1/n$; therefore, an inaccuracy in the density calculation that causes $n \leq 0$ even at a single spatial point immediately breaks down the whole calculation. Similarly, the calculation of $\delta T_{\mathrm{TF}}/\delta n$ also induces challenges because it includes the cube root of $n$, which has multiple branches for $n < 0$. Thus, it is considerably difficult to realize a numerical calculation by implementing Eqs. (6–8) with an update of the electron density.

**B. QHT solved using ESE**

In 1926, Madelung proposed that the Schrödinger equation for a single-particle problem can be transformed into the form of hydrodynamic equations [47]. In 1988, an idea to overcome the above difficulty of accurately calculating the density in the fluid dynamical approach was presented in the field of nuclear physics utilizing Madelung's correspondence, namely, converting the fluid dynamical equation into an equivalent Schrödinger equation [48]. We apply this idea to QHT−TF$\lambda$vW, and Eqs. (6–8) are equivalent to the following ESE:

$$i\hbar\sqrt{\lambda}\frac{\partial\Psi}{\partial t} = \left[\frac{\left\{-i\hbar\sqrt{\lambda}\nabla + e(\mathbf{A} + \mathbf{A}_{XC})\right\}^2}{2m} - e\phi + \frac{\delta E_{XC}}{\delta n} + \frac{\delta T_{TF}}{\delta n}\right]\Psi, \qquad (9)$$

where $\Psi(\mathbf{r}, t)$ is the effective wave function. The electron density $n(\mathbf{r}, t)$ and electric current density $\mathbf{J}(\mathbf{r}, t)$ of Eqs. (2–3) are then given using $\Psi(\mathbf{r}, t)$ as:

$$n = |\Psi|^2, \qquad (10)$$

$$\mathbf{J} = -\frac{e}{m}\text{Re}\left[\Psi^*\left\{-i\hbar\sqrt{\lambda}\nabla + e(\mathbf{A} + \mathbf{A}_{XC})\right\}\Psi\right]. \qquad (11)$$

Thus, a positive definite density can be obtained after the wave function $\Psi(\mathbf{r}, t)$ is obtained by solving Eq. (9). Accurate methods have been developed to solve the Schrödinger equation, even in a spatial region where the density is extremely low; therefore, we may overcome the numerical difficulty of the accurate calculation of the density. It is noteworthy that the term involving $\delta T_w/\delta n$ is transformed into the KE operator in the ESE in Eq. (9). This means that it is unnecessary to calculate quantities that include the $1/n$ term.

Previously, the ESE was used to calculate the ground state in the QHT [17]. However, to the best of our knowledge, the ESE has never been applied to the time-dependent electronic dynamics described by QHT $-$TF$\lambda\nu$W. We also note that the previous application of the ESE in nuclear physics was limited to dynamics described by a rotation-free velocity field [48]. However, our implementation includes velocity fields with rotation because we treat the Lorentz force caused by a magnetic field.

### C. Optical system interacting with the ESE

In our practical implementation, we solve the following four equations for the electromagnetic fields in the Lorentz gauge instead of Eqs. (4) and (5):

$$\frac{\partial \mathbf{E}}{\partial t} = \frac{1}{\epsilon_0 \mu_0} \nabla \times \mathbf{B} - \frac{1}{\epsilon_0} \mathbf{J}, \tag{12}$$

$$\frac{\partial \mathbf{B}}{\partial t} = -\nabla \times \mathbf{E}, \tag{13}$$

$$\frac{\partial \mathbf{A}}{\partial t} = -\mathbf{E} - \nabla \phi, \tag{14}$$

$$\frac{\partial \phi}{\partial t} = -\frac{1}{\epsilon_0 \mu_0} \nabla \cdot \mathbf{A}. \tag{15}$$

To solve Eqs. (12) and (13), we can make full use of numerical methods that are well developed in computational electromagnetic analysis. Therefore, the proposed QHT scheme simultaneously solves Eqs. (9–15). In the next section, we investigate the problem displayed in Fig. 1 (a), in which pulsed light with a linear polarization irradiates a nanosphere described by the JM. For this problem, the calculation proceeds as follows.

The propagation of electromagnetic fields $\mathbf{E}$ and $\mathbf{B}$, including incident pulsed light, is described by Maxwell's equations (12) and (13). The electron dynamics described by the effective wave function $\Psi$ is updated using Eq. (9) with the electromagnetic potentials $\mathbf{A}$ and $\phi$, which are given by Eqs. (14) and (15). When the incident pulsed light that is emitted from the electric current source (see Appendix) reaches the nanosphere and excites the electrons, the excited electrons generate an electric current density $\mathbf{J}$, as defined by Eq. (11). This, in turn, goes into Eq. (12) of Maxwell's equations as a source term. The feedback from the excited electrons to the optical system generates a new electromagnetic field that interacts with the electronic system again and acts as the origin of the optical near field and plasmon resonance.

The numerical formulas to solve Eqs. (9–15) in discretized real-time and -space grids are described in detail in the Appendix.

## III. ILLUSTRATIVE CALCULATIONS

In the previous section, we explained the proposed numerical scheme used to stably execute the time evolution calculations of electric current density and electromagnetic fields based on QHT−TF$\lambda$vW using the ESE. In this section, we present illustrative calculations of the linear and third-order nonlinear optical responses of a metallic nanosphere using this method. The results calculated based on QHT−TF$\lambda$vW using the ESE are compared with those obtained using TDDFT with the JM, which reveals the reliability and limitations of the QHT.

### A. Setup for the calculation of a nanosphere

Figure 1 (a) displays the studied system, in which a metallic nanosphere with a diameter $a$ is subjected to incident light that is specified by the electromagnetic fields $\mathbf{E}^{(i)}$ and $\mathbf{B}^{(i)}$. The nanosphere is described by the JM, which replaces an ionic structure with a positive background density $n_{\mathrm{JM}}^{(+)}(\mathbf{r})$ of a spherical shape with the following shape boundary [2]:

$$n_{\mathrm{JM}}^{(+)} = \begin{cases} n_s & \text{if} \quad |\mathbf{r}| \leq a, \\ 0 & \text{if} \quad |\mathbf{r}| > a, \end{cases} \tag{16}$$

where $n_s = \left((4\pi)r_s^3/3\right)^{-1}$ with the Wigner-Seitz radius $r_s$ of the medium. We selected a value of $r_s = 3.99$, which corresponds to Na metal. This leads to a diameter of $a = 4.3$ nm and an electron number of $N_e = 1074$. Although this JM is a simple description, TDDFT calculations with the JM may well reproduce the basic features of the optical properties of nanoparticles that manifest typical quantum mechanical effects, such as spatial nonlocality, electron spill-out, and tunneling in metallic nanostructures [2, 26-33], with moderate computational cost. Adiabatic and local density approximations are used for the exchange-correlation potential $\delta E_{\mathrm{XC}}/\delta n$ [49]. The previous study that derived the QHT from TDDFT considered the XC vector potential $\mathbf{A}_{\mathrm{XC}}(\mathbf{r}, t)$ [18]; however, we did not include the effect here for simplicity.

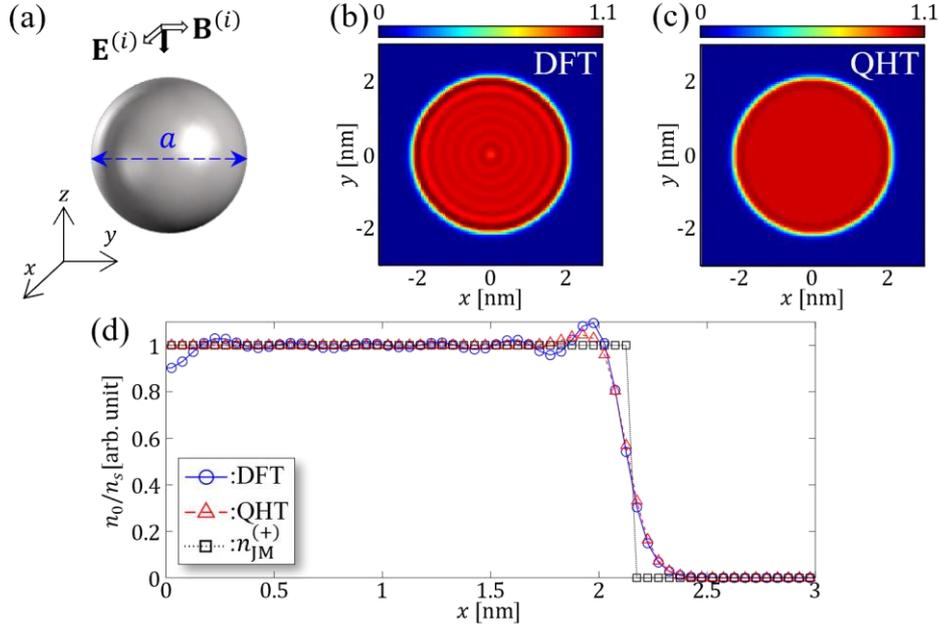

FIG. 1. (a) Schematic of the studied Na nanosphere with the diameter $a$. $\mathbf{E}^{(i)}$ and $\mathbf{B}^{(i)}$ are the electromagnetic fields of the incident light propagating toward the negative $z$ direction. The center of the nanosphere is located at the origin. Electron density distribution in the ground states calculated using the (b) DFT and (c) QHT on the $xy$ plane including the origin. Their amplitudes are normalized by $n_s$. (d) Comparison of the normalized electron densities $n_0/n_s$ along the $x$-axis. The blue solid line with circles (red broken line with triangles) shows the density of the DFT (QHT), and the black dotted line with squares denotes the background positive charge density of the JM, $n_{\mathrm{JM}}^{(+)}$. The phenomenological parameter $\lambda$ in QHT is set to $1/2$.

## B. Ground state

Before exploring the optical responses, we calculate the initial electron density $n_0(\mathbf{r})$ in the ground state, and we use the ESE for the ground state calculation. The time-independent ESE for the effective wave function $\Psi_0(\mathbf{r})$ in QHT−TF$\lambda$vW is given by:

$$\eta \Psi_0 = \left[ -\frac{\hbar^2}{2m}\sqrt{\lambda}\nabla^2 + \frac{1}{\sqrt{\lambda}}\left( -e\phi_0 + \frac{\delta E_{\mathrm{XC}}}{\delta n} + \frac{\delta T_{\mathrm{TF}}}{\delta n} \right) \right] \Psi_0, \tag{17}$$

where $\eta$ is the energy eigenvalue of the ground state. The static scalar potential $\phi_0(\mathbf{r})$ satisfies the Poisson equation:

$$\nabla^2 \phi_0 = -\frac{e\left(n_{\text{JM}}^{(+)} - n_0\right)}{\epsilon_0}. \tag{18}$$

The electron density in the ground state is obtained from the wave function by $n_0(\mathbf{r}) = |\Psi_0(\mathbf{r})|^2$.

To solve Eqs. (17) and (18) numerically, we discretize a spatial area using a three-dimensional Cartesian grid. The spatial grid spacings $\Delta x$, $\Delta y$, and $\Delta z$ are set to 0.05 nm. This value is chosen so that the electron spill-out at the subnanometer scale can be accurately described, as will be demonstrated later. The computational domain is set to a 14.6 nm cube, which is substantially larger than the diameter of the nanosphere $a$. A finite difference formula is used for the spatial derivatives, and an imaginary time propagation method [50] is used to calculate the ground state self-consistently. These methods are explained in the Appendix.

Figure 1 (b) and (c) display the spatial distribution of the ground-state electron density $n_0$ calculated using DFT and QHT, respectively. The phenomenological parameter $\lambda$ in Eq. (6) is set to $\lambda = 1/2$. Figure 1 (c) exhibits a lack of ring-like patterns. These patterns originate from the Friedel oscillation, which is characterized by the Fermi wavenumber, and the quantum mechanical shell effect formed by the electronic orbitals $\psi_j$ interfered with each other. Therefore, QHT, which is categorized as a semiclassical approach without the orbitals, cannot reproduce the density oscillation. For a clearer comparison, we show their electron density distributions along the $x$-axis in Fig. 1 (d). The blue broken line with circles and the red broken line with triangles display the electron density of the DFT and QHT, respectively. As a reference, we also plotted the background positive density of the JM, $n_{\text{JM}}^{(+)}$, using a black dotted line with squares.

Although QHT does not reproduce the oscillation caused by the quantum effect that appears at $x \leq 2$ nm, the density tail at $x > 2$ nm, including the electron spill-out, exhibits a good agreement between QHT and DFT. This agreement is due to an appropriate choice of the $\lambda$ value, which critically affects the density tail and electron spill-out. In the following subsections, we use $\lambda = 1/2$ in the real-time calculations for linear and nonlinear optical responses.

## C. Linear response

In this subsection, we numerically solve the QHT using the ESE, given in Eqs. (9–11), coupled with the electromagnetic fields described by Eqs. (12–15) employing real-time and -space grids. The linear response of the metallic nanosphere shown in Fig. 1 is calculated using the dipole approximation.

As stated in previous studies [14-19, 46], QHT cannot describe the damping effect that originates from electron-electron interactions among the electronic orbitals $\psi_j$. Therefore, a friction term to model the damping effect is added to Eq. (6) with the phenomenological damping rate parameter $\gamma$. However, in QHT using the ESE, it is difficult to introduce such a friction term in the Schrödinger equation. Thus, instead of the friction term, we introduce a background lossy medium that adds the following conductive current density $\mathbf{J}_C(\mathbf{r}, t)$ in Eq. (12):

$$\mathbf{J}_C(\mathbf{r}, t) = \sigma g(\mathbf{r})[\mathbf{E}(\mathbf{r}, t) - \mathbf{E}(\mathbf{r}, t = 0)], \tag{19}$$

where $\sigma$ is a phenomenological conductivity parameter, and $g(\mathbf{r})$ specifies the spatial distribution. The second term in the square brackets in Eq. (19) is added such that the conductive current vanishes, $\mathbf{J}_C = 0$, in the ground state before light irradiation. Among the possible candidates for $g$, we model it in the present study as:

$$g(\mathbf{r}) = \frac{n_0(\mathbf{r})}{n_s}. \tag{20}$$

The source term $\mathbf{J}$ in Maxwell's equation in Eq. (12) is then modified as:

$$\mathbf{J}(\mathbf{r}, t) = \mathbf{J}_Q(\mathbf{r}, t) + \mathbf{J}_C(\mathbf{r}, t), \tag{21}$$

where $\mathbf{J}_Q$ is the electric current density of the QHT, defined by Eq. (11).

We adopt a dipole approximation in the linear response calculation because the diameter of the sphere is sufficiently smaller than the light wavelength. As a perturbation, we use the following temporarily impulsive and spatially uniform field with polarization in the $x$-direction:

$$\mathbf{E}^{(i)}(t) = -E_d \delta(t) \hat{\mathbf{x}}, \tag{22}$$

where $\delta$ is the delta function, and $E_d$ denotes the amplitude. $\hat{\mathbf{x}}$ is the unit vector along the $x$-axis. We describe this impulsive electric field using the vector potential $\mathbf{A}(\mathbf{r}, t)$ in Eq. (14). Integrating Eq. (14) over an infinitesimally short time interval from $0_-$ to $0_+$, the vector potential $\mathbf{A}$ shows an abrupt change from $\mathbf{A}(\mathbf{r}, t) = 0$ for $t < 0_-$ to

$$\mathbf{A}(\mathbf{r}, t = 0_+) = -\int_{0_-}^{0_+} dt \, [-E_d \delta(t) \hat{\mathbf{x}} + \nabla \phi] = E_d \hat{\mathbf{x}}. \tag{23}$$

The impulsive electric field in Eq. (22) includes the entire spectral domain. Therefore, we can calculate the linear optical response of the entire spectral domain simultaneously utilizing the vector potential of Eq. (23) as the initial condition. The other initial conditions of the QHT in the linear response simulation are as follows: $\Psi = \Psi_0$, $n = n_0$, $\phi = \phi_0$, $\mathbf{E} = -\nabla \phi_0$, and $\mathbf{J} = \mathbf{B} = 0$.

The linear optical absorption cross-section is given by

$$S_{\text{abs}}(\omega) = \sqrt{\frac{\mu_0}{\epsilon_0}}\,\omega\text{Im}[\alpha(\omega)], \tag{24}$$

where $\alpha$ denotes the polarizability of the nanosphere in the $x$-direction. This is calculated using the following expressions for the total dipole moment $\mathbf{d}$, which includes both $\mathbf{J}_Q$ and $\mathbf{J}_C$:

$$\alpha(\omega) = -\frac{1}{E_d}\int_0^{T_{\max}} dt\,\left[w(t)d_x(t)e^{i\omega t}\right], \tag{25}$$

$$\mathbf{d}(t) = \int_0^t dt\left[\iiint dxdydz\,\{\mathbf{J}_Q(\mathbf{r},t) + \mathbf{J}_C(\mathbf{r},t)\}\right]. \tag{26}$$

$d_x$ is the $x$-component of $\mathbf{d}$, and $w(t)$ is the window function that is used to erase spurious oscillations originating from the finite period of the Fourier transformation:

$$w(t) = 1 - \frac{1}{3}\left(\frac{t}{T_{\max}}\right)^2 + 2\left(\frac{t}{T_{\max}}\right)^3, \tag{27}$$

where $T_{\max}$ is the time evolution calculation duration. In the present study, $T_{\max}$ is set to 110 fs, and the temporal grid spacing $\Delta t$ is set to $9.628 \times 10^{-5}$ fs. Using this value, the stability condition is satisfied for Maxwell's equations (details are described in Appendix). $E_d$ of Eq. (22) is chosen to be 0.41 MV/m ($8 \times 10^{-7}$ a.u.).

We compare the results of the QHT and TDDFT. The TDDFT calculation method is described below. We employ the length gauge in the dipole approximation. Namely, while Eq. (2) remains unchanged, Eqs. (1, 3–4) are modified as follows:

$$i\hbar\frac{\partial\psi_j}{\partial t} = \left[-\frac{\hbar^2}{2m}\nabla^2 - e\phi + \frac{\delta E_{\text{XC}}}{\delta n} + V_{\text{ext}}\right]\psi_j, \tag{28}$$

$$\mathbf{J} = -\frac{e}{m} \mathrm{Re} \left[ \sum_{j}^{\mathrm{occ}} \psi_j^* (-i\hbar\nabla) \, \psi_j \right], \tag{29}$$

$$\nabla^2 \phi = -\frac{e\left(n_{\mathrm{JM}}^{(+)} - n\right)}{\epsilon_0}, \tag{30}$$

where the incident light is treated by $V_{\mathrm{ext}}(\mathbf{r}, t) = e\mathbf{E}^{(i)}(t) \cdot \mathbf{r}$. We use the impulsive electric field of Eq. (22) for the linear response calculation, which results in the contribution of the initial phase of $\psi_j$ in Eq. (28). All calculations were performed using SALOMN, which is an open-source code developed by our group [51].

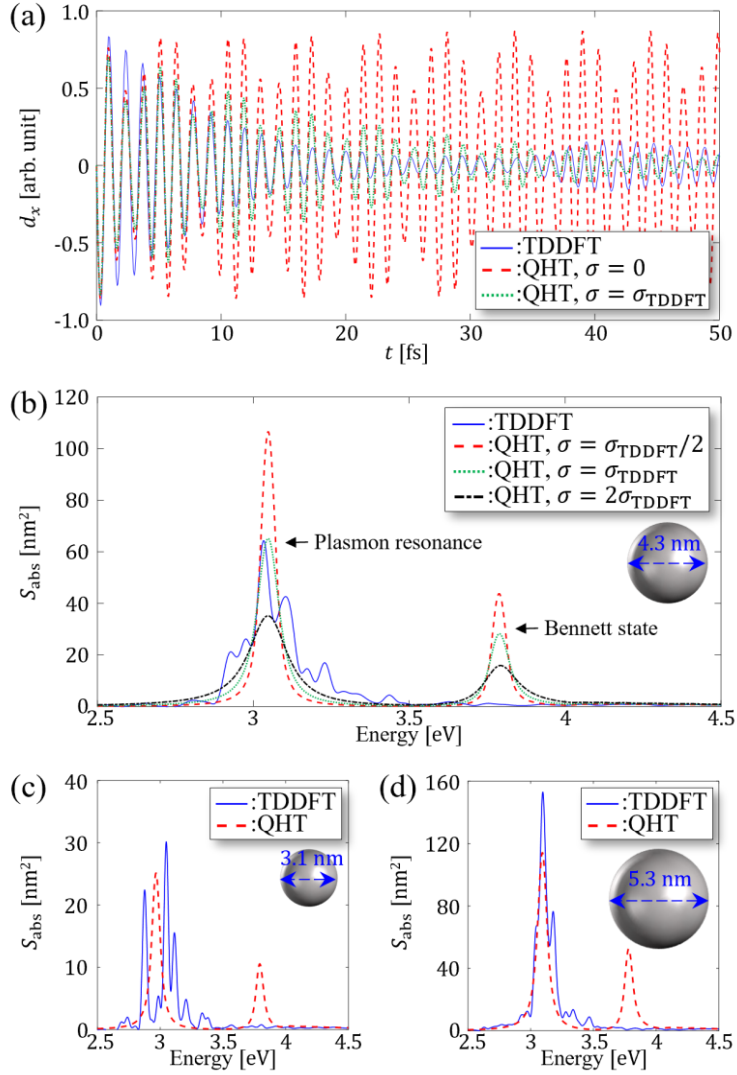

FIG. 2. (a) Temporal profile of the total dipole moment $d_x$ (the $x$-component of $\mathbf{d}$) for a nanosphere with a diameter of $a = 4.3$ nm subjected to the impulsive incident field of Eq. (22). The blue solid line shows the calculation using TDDFT, and the red broken and green dotted lines show the QHT using the ESE with $\sigma = 0$ and $\sigma_{\text{TDDFT}}$, respectively. (b) Spectral distribution of the linear optical absorption cross-section $S_{\text{abs}}$ for the same nanosphere. The blue solid line shows the TDDFT, and the other three lines (red, green, and black) show the QHT with $\sigma = \sigma_{\text{TDDFT}}/2$, $\sigma_{\text{TDDFT}}$, and $2\sigma_{\text{TDDFT}}$, respectively. $S_{\text{abs}}$ of two different nanospheres with (c) $a = 3.1$ nm and (d) $a = 5.3$ nm with $\sigma = \sigma_{\text{TDDFT}}$. The blue and red lines show the results using the TDDFT and QHT, respectively.

Figure 2 (a) shows the temporal profile of the total dipole moment $d_x$ (the $x$-component of $\mathbf{d}$) for the nanosphere shown in Fig. 1, with a diameter of $a = 4.3$ nm and an electron number of $N_e = 1074$, subjected to the impulsive incident field $\mathbf{E}^{(i)}$ of Eq. (22). In the TDDFT calculation exhibited by the blue solid line, the damping effect that originates from electron-electron interactions among the electronic orbitals $\psi_j$ appears. The dipole moment by the TDDFT induces damping by $t \approx 30$ fs, after which a small oscillation continues. The red broken line is the result of the QHT using the ESE with $\sigma = 0$. As described above, QHT with no loss does not include the damping effect, which results in an undamped oscillation. To reproduce damping in the TDDFT calculation, we choose the value of $\sigma$ to be $\sigma_{\text{TDDFT}} = 5.05 \times 10^3$ S/m ($1.10 \times 10^{-3}$ a.u.). The green dotted line in Fig. 2 (a) shows the result obtained using QHT with this value of $\sigma_{\text{TDDFT}}$. Although this value of $\sigma$ is determined using a comparison with TDDFT in the present study, it may also be determined by referring to the experimentally observed absorption spectrum of nanoparticles.

Figure 2 (b) shows the spectral distribution of the linear optical absorption cross-section $S_{\text{abs}}$. The blue solid and red broken lines are the TDDFT and QHT results with $\sigma_{\text{TDDFT}}$. Their maximum peaks at approximately 3 eV originate from the plasmon resonance, the amplitudes and frequencies of which exhibit good agreement between the TDDFT and QHT. However, TDDFT includes several peaks around the main peak that are not observed in QHT. These additional peaks are created from a number of electronic orbitals $\psi_j$. QHT cannot reproduce such peaks, which indicates its limitations. Furthermore, we observe another peak at approximately 3.8 eV only in QHT. This peak is known as the Bennett state, and it is caused by the nonuniformity of the electron density [52]. This state is known to also appear in TDDFT as a small peak or shoulder [30]; however, QHT overestimates its appearance. Recent studies have revealed that this overestimation of the Bennett state can be mitigated using more sophisticated QHTs beyond the TF$\lambda v$W level, such as dynamical correction or high-order KE functionals [18, 46], which are not considered in the present study. The red broken and green chained lines in Fig. 2 (b) exhibit spectra with different choices of $\sigma$: $\sigma_{\text{TDDFT}}/2$ and $2\sigma_{\text{TDDFT}}$. These lines clearly show the importance of the choice of $\sigma$ because it controls the peak widths of the plasmon resonance and Bennett state while maintaining their resonant energies.

To illustrate the robustness of QHT using the ESE, we calculate nanospheres of different sizes: (i) $a = 3.1$ nm ($N_e = 398$); (ii) $a = 5.3$ nm ($N_e = 2030$). Figure 2 (c) and (d) display their $S_{\text{abs}}$ for $a = 3.1$ and 5.3 nm, respectively, where $\sigma$ is set to $\sigma_{\text{TDDFT}}$. In Fig. 2 (c), for the $a = 3.1$ nm nanosphere, the TDDFT calculation plotted using the blue solid line shows several fragmented peaks around the plasmon resonance frequency of 3 eV, owing to a decrease in the number of occupied orbitals $\psi_j$ and the larger energy spacing between them. The QHT plotted using the red broken line cannot reproduce such spectra. In Fig. 2 (d), for the $a = 5.3$ nm nanosphere, a good agreement is observed between the TDDFT and QHT results. In particular, the spectral widths of the plasmon resonance coincide well with each other, compared with the case of the $a = 4.3$ nm nanosphere, shown in Fig. 2 (b). The plasmon in the TDDFT calculation is almost single-peaked, owing to an increase in the number of orbitals, which makes the energy levels more dense.

### D. Nonlinear response

We now consider the calculation of the nonlinear optical response of the $a = 4.3$ nm nanosphere displayed in Fig. 1. The following incident light pulse $\mathbf{E}^{(i)}$ polarized along the $x$-axis is employed:

$$\mathbf{E}^{(i)}(t) = F \cos^2\left[\frac{\pi}{T}\left(t - \frac{T}{2}\right)\right]\sin(\omega_i t)\,\hat{\mathbf{x}} \quad (0 < t < T), \tag{31}$$

where $F$, $T$, and $\omega_i$ are the amplitude, pulse duration, and fundamental frequency of the pulse, respectively. In this study, $F$ is set to $274$ MV/m ($5.34 \times 10^{-4}$ a.u.), which corresponds to the pulse intensity $I = 10^{10}$ W/cm$^2$, and $T$ is fixed at $55$ fs. In the QHT calculation, this incident pulse is emitted from an electric current source, as described in the Appendix. The initial conditions and other computational details of the QHT are the same as those in the previous subsection, with the exception of $\mathbf{A}(\mathbf{r}, t = 0) = 0$.

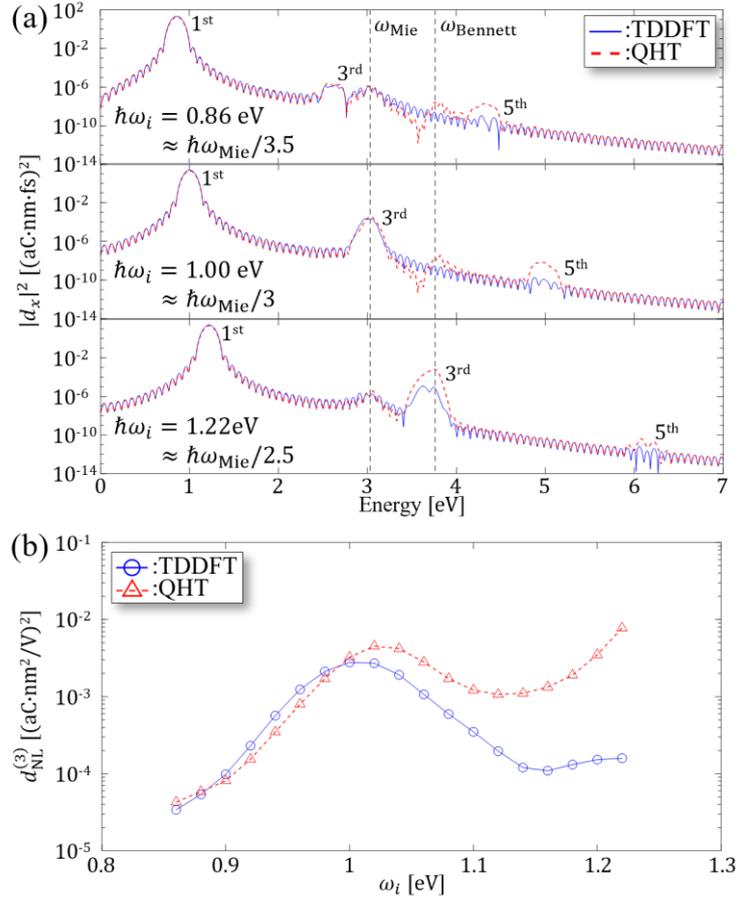

FIG. 3. (a) Power spectra of the total dipole moment $d_x$ for the nanosphere with a diameter of $a = 4.3$ nm. The nanosphere is subjected to the pulsed incident light described in Eq. (31). The blue solid (red broken) lines are from the TDDFT (QHT) calculations. The two vertical gray dashed lines indicate the frequencies of the plasmon resonance $\omega_{\mathrm{Mie}}$ and Bennett state $\omega_{\mathrm{Bennett}}$. The top, middle, and bottom panels correspond to the pulse irradiation with $\hbar\omega_i = 0.86 \ (\approx \hbar\omega_{\mathrm{Mie}}/3.5)$, $1.00 \ (\approx \hbar\omega_{\mathrm{Mie}}/3)$, and $1.22 \ (\approx \hbar\omega_{\mathrm{Mie}}/2.5)$ eV, respectively. (b) $\omega_i$ dependence of the third-order optical nonlinearity $d_{\mathrm{NL}}^{(3)}$ defined by Eq. (32). The blue solid line with circles (red broken line with squares) denotes the results obtained using TDDFT (QHT).

Figure 3 (a) shows the power spectra of the total dipole moment $d_x$, where the Fourier transformation is applied from $t = 0$ to $T_{\mathrm{max}}$ with the window function $w(t)$ defined in Eq. (27). The blue solid and red broken lines are calculated by the TDDFT and QHT using the ESE, respectively. The two vertical gray dashed lines indicate the frequencies of the plasmon resonance $\omega_{\mathrm{Mie}}$ and Bennett state $\omega_{\mathrm{Bennett}}$. We show the calculations for three incident pulses with the following fundamental frequencies: $\hbar\omega_i =$

0.86 ($\approx \omega_{\text{Mie}}/3.5$), 1.00 ($\approx \hbar\omega_{\text{Mie}}/3$), and 1.22 ($\approx \hbar\omega_{\text{Mie}}/2.5$) eV. The top panel of Fig. 3 (a) corresponds to the pulse irradiation with $\hbar\omega_i = 0.86$ eV. Below 3.5 eV, we observe three peaks: the fundamental and 3$^{\text{rd}}$ harmonic generation at approximately 0.86 and 2.58 eV, respectively, and the plasmon resonance at $\omega_{\text{Mie}}$. QHT succeeded in reproducing these peaks. However, in the energy region higher than 3.5 eV, two additional peaks appear in QHT, but they are not found in TDDFT. The first peak at approximately 3.8 eV is owing to the Bennett state, which is overestimated in QHT, as demonstrated in Fig 2 (b). The second peak at approximately 4.3 eV corresponds to the 5$^{\text{th}}$ harmonic generation. The 3$^{\text{rd}}$ harmonic generation of QHT and TDDFT coincide accurately with each other; however, the 5$^{\text{th}}$ harmonic generation is only visible in the QHT. The reason for this is currently unclear. The Bennett state that appears strongly in the QHT may affect this phenomenon. The middle panel of Fig. 3 (a) corresponds to the pulse irradiation with $\hbar\omega_i = 1.00$ eV. The 3$^{\text{rd}}$ harmonic generation of $\omega_i$ overlaps with the plasmon resonance $\omega_r$. Therefore, the 3$^{\text{rd}}$ nonlinear signal is plasmonically enhanced here, and the QHT reproduces the enhanced nonlinearity well. In the case of the top panel, $\hbar\omega_i = 0.86$ eV, two peaks are formed by the Bennett state and 5$^{\text{th}}$ harmonic generation. Similarly, the 5$^{\text{th}}$ harmonic generation is observed only in the QHT. The bottom panel corresponds to the pulse irradiation with $\hbar\omega_i = 1.22$ eV. In this case, the 3$^{\text{rd}}$ harmonic generation overlaps with the Bennett state. A large difference appears in the 3$^{\text{rd}}$ nonlinear signal between the TDDFT and QHT because the Bennett state appears strongly in QHT. The 5$^{\text{th}}$ harmonic generation is slightly visible at approximately 6 eV in the QHT.

Finally, we define the following quantity to explore the difference in third-order nonlinearity between the TDDFT and QHT more quantitatively:

$$d_{NL}^{(3)} = \frac{\int_{2.5\omega_i}^{3.5\omega_i} d\omega \, [|d_x(\omega)|^2]}{\int_0^\infty d\omega \left[\left|E_x^{(i)}(\omega)\right|^2\right]},$$ (32)

where $d_x(\omega)$ and $E_x^{(i)}(\omega)$ denote the Fourier transformation of the $x$-component of the quantities given by Eqs. (26) and (31), respectively, with the window function $w(t)$ in Eq. (27). The value of $d_{NL}^{(3)}$ indicates the magnitude of the 3$^{rd}$ harmonic generation for an incident pulse with $\omega_i$. Figure 3 (b) shows the calculated $d_{NL}^{(3)}$. The blue solid line with the circles and the red broken line with the triangles represent the results obtained using TDDFT and QHT, respectively. The two lines exhibit their peaks at approximately $\hbar\omega_i = 1$ eV, which reflects the plasmonically enhanced nonlinearity, as demonstrated the middle panel of Fig. 3 (a). The lines exhibit a reasonable agreement below $\hbar\omega_i = 1.1$ eV. Therefore, we conclude that QHT is capable of describing third-order nonlinearity with reasonable accuracy. However, above that frequency, the value of the QHT $d_{NL}^{(3)}$ increases as frequency increases, and the discrepancy between the two theories increases. This is owing to the Bennett state, and it reflects the limitation of the present approach based on QHT−TF$\lambda$vW. More sophisticated QHTs that rely on dynamical correction or high-order KE functionals are required to mitigate the discrepancy in the future.

## IV. CONCLUSION

This study proposes a numerical scheme that enables stable calculations of linear and nonlinear electron dynamics induced by an optical field described using QHT−TF$\lambda$vW. The key component of our scheme is to rewrite the conventional equations of the QHT−TF$\lambda$vW to the ESE, and to solve the ESE coupled with Maxwell's equations using real-time and -space grids. In QHT described by the ESE, all numerical instabilities originating from $1/n$ and the cube root of $n$ terms that appear in KE functionals have been removed, thereby making it easy to expand the computational targets to nonlinear optical phenomena without any approximations. We believe that the proposed numerical method of QHT using the ESE will be useful for exploring nanoplasmonics that include quantum mechanical effects.

The result was compared with that of the calculations for the fully quantum mechanical DFT with the JM for the ground state and the TDDFT calculations for the linear and nonlinear optical responses of a metallic nanosphere. We have shown that an appropriate choice of $\lambda$ in the QHT is essential to reproduce the density tail in the ground state and the plasmonic spectrum in the linear response in the (TD)DFT calculations. In particular, our QHT calculation successfully reproduced the third-order nonlinearity enhanced by plasmon resonance in the TDDFT calculation. However, at an energy region higher than the plasmon resonance, the QHT overestimated the Bennett state in the linear response, which originated from the nonuniformity of the electron density. Owing to this overestimation in the linear response, the nonlinear spectrum calculated using QHT gradually deviated from that of the TDDFT at an energy region higher than the plasmon. To remedy this overestimation, more sophisticated QHTs that rely on dynamical correction or high-order KE functionals may be necessary.

**ACKNOWLEDGEMENTS**

The authors thank Prof. T. Nakatsukasa (University of Tsukuba) for his useful advice on the imaginary time propagation method. This research was supported by the JST–CREST under Grant No. JP–MJCR16N5, the JSPS Research Fellowships for Young Scientists, and the JSPS KAKENHI under Grant Nos. 20J00449 and 20H02649.

**APPENDIX**

In this appendix, we present mathematical formulas to be used in the computational codes for the present QHT using the ESE.

**1. Finite-difference method**

Our proposed QHT scheme using the ESE, given in Eqs. (9–11), coupled with the electromagnetic fields described by Eqs. (12–15), includes first- and second-order derivatives with respect to time and space. Using a finite-difference method, we approximate a $\nu$-th order derivative of an arbitrary function $f(x)$ by a finite difference involving $2\mu + 1$ grids with spacing $\Delta x$ as

$$\frac{\partial^\nu f(x)}{\partial x^\nu} \approx \sum_{i=-\mu}^{\mu} \left[ \frac{c_i f(x + i\Delta x)}{\Delta x^\nu} \right], \tag{A1}$$

where $c_i$ denotes the $i$-th coefficient, and $\nu$ and $\mu$ are positive integers satisfying $2\mu \geq \nu$. By applying the Taylor expansion to $f(x + i\Delta x)$ up to the $2\mu$-th order, we obtain $2\mu + 1$ simultaneous equations with respect to $c_i$. The $j$-th equation is given by

$$\frac{1}{(j-1)!} \sum_{i=-\mu}^{\mu} \left[ i^{(j-1)} c_i \right] = \delta_{(j-1)\nu}, \tag{A2}$$

where $\delta_{ij}$ is the Kronecker delta, and $c_i$ is determined by solving Eq. (A2).

## 2. Imaginary time propagation method

An imaginary time propagation method was employed to determine the electronic ground state $\Psi_0$ discussed in Sec. II-B [50]. In the method applied to Eq. (17), $\Psi_0$ is obtained by recursively updating the following equations until $n_0$ and $\eta$ converge:

$$\Psi_0(\mathbf{r}, t + \Delta t) = [1 - \Delta t\{h(\mathbf{r}, t) - \eta(t)\}]\Psi_0(\mathbf{r}, t), \tag{A3}$$

$$h(\mathbf{r}, t) = -\frac{\hbar^2}{2m}\sqrt{\lambda}\nabla^2 + \frac{1}{\sqrt{\lambda}}\big(-e\phi_0(\mathbf{r}, t) + V_{\text{XC}}(\mathbf{r}, t) + V_{\text{TF}}(\mathbf{r}, t)\big), \tag{A4}$$

$$\eta(t) = \int dv\,[\Psi_0^*(\mathbf{r}, t)h(\mathbf{r}, t)\Psi_0(\mathbf{r}, t)], \tag{A5}$$

where $V_{\text{XC}}$ and $V_{\text{TF}}$ are the functional derivatives $\delta E_{\text{XC}}/\delta n$ and $\delta T_{\text{TF}}/\delta n$, respectively. $\Delta t$ is chosen to be $1.21 \times 10^{-3}$ fs, which corresponds to 0.05 a.u.. At each time step, the Hamiltonian is updated using the wave function $\Psi_0(\mathbf{r}, t + \Delta t)$.

The Laplacians that appear in Eq. (A4) and the Poisson equation, Eq. (18), are solved using the finite-difference method with $\mu = 4$. A conjugate gradient method executed in real-space grids is used to solve the Poisson equation.

## 3. Finite-difference time-domain method to solve the QHT using ESE coupled with Maxwell equations

A finite-difference time-domain (FDTD) method was first used to solve Maxwell's equations [53], Eqs. (12) and (13), where $\mathbf{E}$ and $\mathbf{B}$ are recursively updated through each other. Subsequently, the FDTD method is applied to the time-dependent Schrödinger equation, referred to as FDTD-Q [54], where its wave function is separated into real and imaginary parts, which are also recursively updated through each other. In the present study, we employ two FDTD schemes and combine them with a similar scheme to solve the electromagnetic potentials of Eqs. (14) and (15) on real-space and -time grids.

In our program, we neglect $\mathbf{A}_{\text{XC}}$, and the effective wave function $\Psi$ of Eq. (9) is updated as follows:

$$\Psi_r^{t+\Delta t} = \Psi_r^t - \frac{\hbar\sqrt{\lambda}\Delta t}{2m}\nabla^2\Psi_i^{t+\frac{\Delta t}{2}}$$

$$-\frac{e\Delta t}{2m}\left\{\nabla\cdot\left(\mathbf{A}^{t+\frac{\Delta t}{2}}\Psi_r^{t+\frac{\Delta t}{2}}\right) + \mathbf{A}^{t+\frac{\Delta t}{2}}\cdot\left(\nabla\Psi_r^{t+\frac{\Delta t}{2}}\right)\right\}$$

$$+\frac{\Delta t}{\hbar\sqrt{\lambda}}\left(\frac{e^2\left|\mathbf{A}^{t+\frac{\Delta t}{2}}\right|^2}{2m} - e\phi^{t+\frac{\Delta t}{2}} + V_{\text{XC}}^t + V_{\text{TF}}^t\right)\Psi_i^{t+\frac{\Delta t}{2}}, \tag{A6}$$

$$\Psi_i^{t+\Delta t} = \Psi_i^t + \frac{\hbar\sqrt{\lambda}\Delta t}{2m}\nabla^2\Psi_r^{t+\frac{\Delta t}{2}}$$

$$-\frac{e\Delta t}{2m}\left\{\nabla\cdot\left(\mathbf{A}^{t+\frac{\Delta t}{2}}\Psi_i^{t+\frac{\Delta t}{2}}\right) + \mathbf{A}^{t+\frac{\Delta t}{2}}\cdot\left(\nabla\Psi_i^{t+\frac{\Delta t}{2}}\right)\right\}$$

$$-\frac{\Delta t}{\hbar\sqrt{\lambda}}\left(\frac{e^2\left|\mathbf{A}^{t+\frac{\Delta t}{2}}\right|^2}{2m} - e\phi^{t+\frac{\Delta t}{2}} + V_{\text{XC}}^t + V_{\text{TF}}^t\right)\Psi_r^{t+\frac{\Delta t}{2}}, \tag{A7}$$

where $\Psi_r$ and $\Psi_i$ indicate the real and imaginary parts of $\Psi$. The superscripts denote the time coordinate. The space derivatives that appear in Eqs. (A6) and (A7) are evaluated using the finite-difference method with $\mu = 4$, which is also used to execute the electric current density in Eq. (11). In the above algorithm, the times of $V_{\text{XC}}$ and $V_{\text{TF}}$ are taken as $t$, not $t + \Delta t/2$, because an accumulation of numerical errors appears if we choose time $t + \Delta t/2$. We thoroughly checked that the choice of time $t$ in $V_{\text{XC}}$ and $V_{\text{TF}}$ does not cause inaccuracies by examining the convergence of the calculation as the time step changes.

Subsequently, the electromagnetic fields are updated using Eqs. (12) and (13):

$$\mathbf{E}^{t+\Delta t} = \mathbf{E}^t + \frac{\Delta t}{\epsilon_0 \mu_0} \nabla \times \mathbf{B}^{t+\frac{\Delta t}{2}} - \frac{\Delta t}{\epsilon_0} \left( \mathbf{J}_Q^{t+\frac{\Delta t}{2}} + \mathbf{J}_C^t + \mathbf{J}_S^{t+\frac{\Delta t}{2}} \right), \tag{A8}$$

$$\mathbf{B}^{t+\Delta t} = \mathbf{B}^t + \Delta t \nabla \times \mathbf{E}^{t+\frac{\Delta t}{2}}. \tag{A9}$$

For the same reason given for $V_{\mathrm{XC}}$ and $V_{\mathrm{TF}}$, the time coordinate of $\mathbf{J}_C$ is chosen as $t$. $\mathbf{J}_S$ is an electric current source that emits an incident pulse with a finite width. In Sec. III-D, $\mathbf{J}_S$ is set to be

$$\mathbf{J}_S = -\frac{2\delta_{kk_S}}{\Delta z} \sqrt{\frac{\mu_0}{\epsilon_0}} \mathbf{E}^{(i)}, \tag{A10}$$

where $\mathbf{E}^{(i)}$ is the temporal profile of the pulse, defined in Eq. (31). $k$ is the index of the space grid along the $z$-axis, and $k_S$ is the index along the $z$-axis, where $\mathbf{J}_S$ is located. This $\mathbf{J}_S$ is uniformly placed on the $xy$-plane, and it emits a planar pulse.

Finally, the electromagnetic potentials in Eqs. (14) and (15) are updated as follows:

$$\mathbf{A}^{t+\Delta t} = \mathbf{A}^t - \Delta t \left( \mathbf{E}^{t+\frac{\Delta t}{2}} + \nabla \phi^{t+\frac{\Delta t}{2}} \right), \tag{A11}$$

$$\phi^{t+\Delta t} = \phi^t - \frac{\Delta t}{\epsilon_0 \mu_0} \nabla \cdot \mathbf{A}^{t+\frac{\Delta t}{2}}. \tag{A12}$$

For the spatial derivatives that appear in Eqs. (A8) and (A9) and (A11) and (A12), we applied the finite-difference method with $\mu = 1$. Then, the choice of temporal spacing $\Delta t$ is restricted by the Courant-Friedrichs-Lewy condition [53]:

$$\Delta t < \sqrt{\frac{\epsilon_0 \mu_0}{\left(\frac{1}{\Delta x}\right)^2 + \left(\frac{1}{\Delta y}\right)^2 + \left(\frac{1}{\Delta z}\right)^2}}, \tag{A13}$$

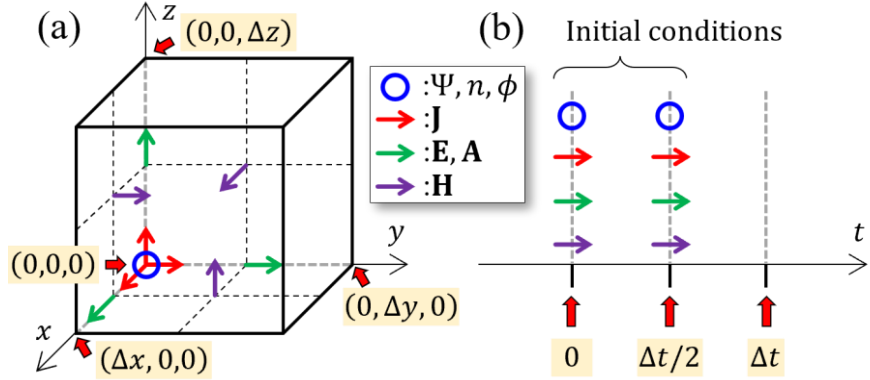

FIG. 4. (a) Space and (b) time configurations of the physical quantities on the spatiotemporal grid in the QHT using the ESE. The blue circles denote scalar quantities, $\Psi$, $n$, and $\phi$, and the red, green, and purple arrows are the vector quantities, $\mathbf{J}$, $\mathbf{E}$, $\mathbf{A}$, and $\mathbf{H}$, respectively.

These physical quantities are arranged into spatiotemporal grids, as shown in Figure 4. The space configuration is displayed in Fig. 4 (a), where the scalar physical quantities are represented by the blue circles; $\Psi$, $n$, and $\phi$ are arranged at the vertex; and vector quantities $\mathbf{J}$, $\mathbf{E}$, $\mathbf{A}$, and $\mathbf{H}$ are at the vertex, edge, and face. If a particular quantity is required at different positions (for example, in Eqs. (A6) and (A7), when $\Psi_r$ and $\Psi_i$ on the vertex are updated, $\mathbf{A}$ on the edge is required), then a second-order averaging, $f\left(x + \frac{\Delta x}{2}\right) \approx \frac{f(x+\Delta x)+f(x)}{2}$, is used. Figure 4 (b) shows the time configuration. As this figure indicates, in our simulation, two initial conditions at $t = 0$ and $\Delta t/2$ are required. Using these conditions, we updated all physical quantities. $\Psi = \Psi_0 e^{-i\frac{n}{\hbar}t}$ is used to prepare $\Psi$ at $t = \Delta t/2$.

## 4. Equations of QHT using ESE in Gaussian units

For convenience, we summarize the equations of the QHT using the ESE, Eqs. (9–15), in the Gauss unit system:

$$i\hbar\sqrt{\lambda}\frac{\partial \Psi}{\partial t} = \left[\frac{\left\{-i\hbar\sqrt{\lambda}\nabla + \frac{e(\mathbf{A}+\mathbf{A}_{\mathrm{XC}})}{c}\right\}^2}{2m} - e\phi + \frac{\delta E_{\mathrm{XC}}}{\delta n} + \frac{\delta T_{\mathrm{TF}}}{\delta n}\right]\Psi, \qquad (A14)$$

$$n = |\Psi|^2, \qquad (A15)$$

$$\mathbf{J} = -\frac{e}{m}\,\mathrm{Re}\left[\Psi^*\left\{-i\hbar\sqrt{\lambda}\nabla + \frac{e(\mathbf{A}+\mathbf{A}_{\mathrm{XC}})}{c}\right\}\Psi\right]. \qquad (A16)$$

$$\frac{\partial \mathbf{E}}{\partial t} = c\nabla \times \mathbf{B} - 4\pi\mathbf{J}, \qquad (A17)$$

$$\frac{\partial \mathbf{B}}{\partial t} = -c\nabla \times \mathbf{E}, \qquad (A18)$$

$$\frac{\partial \mathbf{A}}{\partial t} = -c(\mathbf{E} + \nabla\phi), \qquad (A19)$$

$$\frac{\partial \phi}{\partial t} = -\frac{1}{c}\nabla \cdot \mathbf{A}, \qquad (A20)$$

where $c$ is the speed of light.